\documentclass[titlepage,twoside,12pt]{article}
\usepackage{amssymb}
\usepackage{amsfonts}
\begin{document}

\pagenumbering{arabic}
\setcounter{page}{1}

\pagenumbering{arabic}

{\LARGE \bf Notes on Quantum Mechanics and Consciousness} \\ \\

{\bf Elem\'{e}r E Rosinger \\
Department of Mathematics \\
University of Pretoria \\
Pretoria, 0002 South Africa \\
e-mail : eerosinger@hotmail.com} \\ \\

{\bf Abstract} \\

There have lately been a variety of attempts to connect, or even explain, if not in fact,
reduce human consciousness to quantum mechanical processes. Such attempts tend to draw a sharp
and fundamental distinction between the role of consciousness in classical mechanics, and on
the other hand, in quantum mechanics, with an insistence on the assumed exceptional character
of the latter. What is strangely missed, however, is the role of human consciousness as such
in the very discovery or creation of both of these physical theories. And this a priori role
is far more important than all the possible a posteriori interplays between consciousness and
the mentioned two theories of physics, interplays which may happen during one or another
specific experiment, measurement, and so on. In this regard it is suggested that the specific
features human consciousness may exhibit during interactions with quantum mechanical systems
may as well have other explanations which do not appear to be less plausible, or less well
founded. \\ \\

{\bf 1. Introduction} \\

Since the literature on the relationship between consciousness and quantum processes is rather
considerable, here for the sake of brevity, we shall refer specifically to one rather typical
sample presented by the recent paper of Stapp. \\

For a start, let us for a moment take one or more steps back, and from the respective
perspective, note a few features which may be relevant not only to the way one deals with the
relationship between consciousness and quantum processes, but possibly as well to the venture
of quantum mechanics itself. \\

A deficiency, which tends to amount to a rather regrettable feature of studies in quantum
mechanics, is manifested in the fact that, subsequent to the times of John von Neumann ended
in the 1950s, the main players prove, intentionally or not, to have a manifest disregard not
only for the latest conquests of pure mathematics, but also of applied mathematics. What pure
mathematics is concerned, this means a near unanimous limitation to concepts introduced and
developed prior to World War II, such as Hilbert spaces, C*-algebras, and the like, and of
course, to the near exclusive use of the much earlier known scalars given by the usual real or
complex numbers. On the other hand, when it comes to applied mathematics, one cannot but note
with regret that by now classical, well established and massively used concepts are not
employed according to their full relevance and power. For instance, we can mention in this
regard linear and finite dimensional control theory, with concepts such as "state-space",
"observability", "controlability", "reachability", "stability", and so on, celebrated ever
since the 1960s with their successes, among others, in the Kalman-Bucy filters. \\

It may appear to be a first natural reaction on the part of those involved in the cutting edge
pursuit of quantum mechanics to disregard much of everything else in science. After all, it
has always been a widely accepted assumption that, according to our best present knowledge and
understanding, the quantum level is underlying all else in nature. Not to mention that quantum
phenomena are so much different from whatever else we have known earlier. \\
However, two objections may nevertheless arise. \\
First, to the extent that in the study of quantum processes one goes outside of them and deals,
for instance, with consciousness, mind, brain, or for that matter, with any other relevantly
related macroscopic phenomenon, one may do well to be familiar with and make use of the
state-of-the-art concepts in these non-quantum realms. \\
Second, scientific thinking can only benefit from what in more simple terms can be called
cross-fertilization of concepts across a number of different ventures in science, often so
different as at first not to appear being related in any relevant manner. \\

Another feature, ever since the 1920s, of the approaches of most of those involved in the
cutting edge pursuit of quantum mechanics has been the relentless insistence on, and rather
enthusiastic highlightening of what appeared as being the extraordinary and completely
unprecedented difference between that theory, and on the other hand, all other earlier known
theories of physics. \\

Such and other possibly questionable features in quantum mechanical studies have quite likely
had their less than fortunate consequences. \\

Here in this paper we shall mean by quantum mechanics the theory of non-relativistic quantum
systems of finitely many particles. \\ \\

{\bf 2. Are States of Quantum Systems of Any Relevance ?} \\

In control theory, or more at large, in general systems theory, any given system $S$ that is
not supposed to be fixed in time is assigned a so called "state-space" $X_S$ which contains
the set of all its different possible relevant manifestations. \\

In von Neumann's first model of quantum mechanics, Rosinger [1, pp. 9-13], such a
"state-space" is given by the vectors $\psi$ of a suitable Hilbert space ${\cal H}$, vectors
which are called "wave functions", and which are supposed to be normalized, that is, satisfy
the condition $||~ \psi ~|| = 1$. Further, in this model there is also the concept of
"observable", and it is given by self-adjoint operators $A : {\cal H} \longrightarrow {\cal
H}$. Here however one should note a certain difference, when compared with the sense control
theory uses the similarly named concept. Indeed, in quantum mechanics by "observable" one
rather means a specific kind of "observation" or measurement of the quantum system, while in
control theory, the term "observable" describes an important property of the system as a whole,
and not of any of its particular observation or measurement. One can recall in this regard the
celebrated theorem of Kalman, according to which a finite dimensional linear system which is
observable will also be stable, if and only if it is controllable. \\
Similarly, the concept of "state-space", and that of "observable", in its quantum specific
sense, are present in von Neumann's second model, Rosinger [1, pp. 9-13]. \\

On the other hand, in spite of such a presence of the concept of "state-space", one finds as a
pillar of the Copenhagen Interpretation statements such as :

\begin{quote}

"The conception of objective reality of the elementary particles has thus evaporated not into
the cloud of some obscure new reality concept but into the transparent clarity of a
mathematics that represents no longer the behaviour of the particles but rather our knowledge
of this behaviour", Heisenberg.

\end{quote}

In other words, the "state-space" is {\it no longer} supposed to incorporate anything at all
related to the set of all different possible relevant manifestations of the quantum system
which is under consideration. Instead, all that the "state-space" does - and is alleged to be
able to do - is to incorporate but the different possible relevant manifestations of our
knowledge of that quantum system. \\
Such a view obviously is a most remarkable - and in science, unprecedented - {\it total
renunciation} of any possible {\it ontological} relation on our part to quantum systems, and
implies the {\it total resignation} to a mere {\it epistemological} connection. \\

Nevertheless, the Copenhagen Interpretation does not deny that quantum system exists as such
as physical entities, and that they may go through a variety of different physically relevant
manifestations, and do so either all on their own, or in conjunction with their observation or
measurement. \\

Those who may find such a {\it total renunciation} + {\it total resignation} hard to accept,
among them Einstein, Schr\"{o}dinger, or Bohm, may simply ask :

\begin{quote}

Are physicists, and for that matter, the theories of physics, ever to deal with the
"state-spaces" of quantum systems in the physical sense of this concept, that is, as it is
understood in all the other branches of physics, namely, ontologically ? \\
And if not, then who else is supposed to do it ? \\
And what kind of other possible scientific theories may be able to do so ?

\end{quote}

The Copenhagen Interpretation, instead of having a constructive approach to such questions,
seems to take a special pride in the mentioned {\it total renunciation} + {\it total
resignation}, and appears to do so under the pretext of, and in direct proportion to the truly
unprecedented novelty in the whole of modern science of such a controversial position. \\
Naturally, scientists evermore and most eagerly aim at bringing forth truly unprecedented
novelties. The Copenhagen Interpretation is obviously not a stranger to such an attitude. What
is strange and unprecedented in science, however, about that interpretation is the eagerness
to pay such a considerable price renunciation and resignation for trying to make sense out of
the remarkable novelties of quantum phenomena. \\ \\

{\bf 3. What Aspect of Conscious Involvement Is More \\ \hspace*{0.5cm} Important ?} \\

As far as we all know, humankind got to know about classical and quantum mechanics not as a
gift from somewhere outside of itself, but as the discovery or creation of some of its own
members. And needless to say, the respective processes of creation or discovery were acts of
human consciousness. \\
By the way, lest we may overlook it, it is useful to recall that classical mechanics is by no
means a closed or exhausted subject. Indeed, issues such as for instance turbulence in fluid
flow, or even the existence of regular enough solutions of the Navier-Stokes equations, are
still widely open and highly nontrivial problems. \\

In view of that one may consider that the involvement of human consciousness in classical or
quantum mechanics not only starts with the discovery or creation of the respective physical
theories, and keeps going on with their further development, but that such an "a priori"
involvement is a far more deep and relevant manifestation of human consciousness within the
realms of physics, than any other "a posteriori" conscious involvement during one or another
specific physical experiment. \\

In a manifest contradistinction to the above, a great accent is often placed in the literature
dealing with the relationship between consciousness and quantum processes on the significant
difference the consciousness of the human observer plays in classical, as opposed to quantum
mechanics. \\
It is for instance stated that in classical mechanics "... all physically described properties
become completely determined by physically described properties alone, with consciousness a
causally inert, or causally superfluous bystander. Correlations between the physically and
psychologically described properties can be described within a classical physics based
framework, but the psychologically described aspects will remain essentially epiphenomenal
byproducts of brain activity", Stapp. \\

When encountering such or similar views, and especially coming from partial or total
supporters of the Copenhagen Interpretation, the following comment is hard to avoid : \\

The fact that classical mechanics, as correctly described in the above citation, is "causally
complete" only shows that at the "a priori" stage of involvement in it of human consciousness
- namely, of those who created or discovered it - a rather {\it perfect} job was accomplished.
Consequently, in the "a posteriori" stages of involvement of consciousness, such as happens
in the process of various experiments, human consciousness can have it so much easier, in
particular, as a mere witness. \\

On the other hand, when it comes to quantum mechanics, if one starts with the mentioned {\it
total renunciation} + {\it total resignation} promulgated by the Copenhagen Interpretation,
and among others, replaces the ontological "state-space" with a mere epistemological one, one
clearly risks various forms "incompleteness" in the respective theory. After all, even the
Copenhagen Interpretation does not go so far as to deny the physical existence of quantum
system as ontological entities which go through a variety of different physically relevant
manifestations. But then, by denying the very possibility of ever reaching in any theory that
physical ontology, the risk is taken for "incompleteness". \\

In this way, one may note that the attempts within the Copenhagen Interpretation which try to
connect human consciousness with the dynamics of quantum systems are to a good extent trying
to support themselves through a self-fabricated entrapment within the epistemic. Namely  \\

\begin{itemize}

\item one first denies the very possibility of an ontological theory,

\item then one ends up with incompleteness, and finally

\item one has a chance to introduce some active, causal role for human consciousness during
certain experimental processes, in an attempt to explain the whole range of dynamics of
quantum systems.

\end{itemize}

{~} \\

{\bf 4. About Relative Clumsiness} \\

When one observes, for instance, the Moon from the Earth and does so with the naked eye, it is
very hard to assume that one can cause by that any relevant disturbance in the motion of the
Moon. In other words, such an experiment performed relating to the dynamics of the Moon can
quite safely be considered as perfectly {\it non-invasive} in terms of that dynamics. \\

On the other hand, as it is well known, when we implement various experiments in genetic
engineering, many of our actions end up being not only invasive, but simply destructive, even
if they were not meant to be so. \\

The difference between the above two situations, both of them rather within the realms of
classical mechanics, can be seen as a simple manifestation of "relative clumsiness" on our
part, having much to do with the significant differences in the relative sizes we humans have,
when compared with the respective objects of our interest. \\

Compared with the typical scales in the quantum realms, the relative difference between them
and those of our human scales are far larger than when we compare ourselves with the realms
of genes. \\

Consequently, it does not appear too far fetched to consider the possibility that various so
far disregarded {\it invasive} effects may accompany many of the measurement and observation
processes which, as usual, bring together in interaction macroscopic devices with quantum
systems. And one of the consequences of such invasiveness may possibly be what goes under the
name of the "collapse of the wave function", which puts an instant end to the free dynamics of
a quantum system as described by the Schr\"{o}dinger equation, and switches it to some other
state. \\ \\

{\bf 5. Is the Quantum World so Boringly Repetitive ?} \\

One of the arguments brought by the Copenhagen Interpretation in support of the impossibility
of an ontological quantum theory is what appears to be the inherent randomness of a variety of
quantum phenomena, such as for instance, the radioactive decay. \\

As we know from a large variety of situations, randomness experienced on a certain level when
dealing with a given physical process may not necessarily be the sign of an inherent essential
aspect of the respective process. Indeed, sometime, like in the case of statistical mechanics,
it may be the effect of an insufficient knowledge on our part of the precise initial state of
the process. Or it may be a deeper ignorance, namely, of important aspects of the  laws of
dynamics of that physical process. \\

In the case of quantum systems, in addition to some consideration of the possible effects of
the above mentioned "clumsiness", one should not disregard the eventuality that, because of
the same reasons of immense discrepancies of the scales involved, we may miss on certain
relevant features of quantum entities, thus exposing ourselves to what appear as random
manifestations when seen on our level. \\

It is indeed most surprising to note the following rather dramatic {\it discrepancy}. \\
On our human scales, and higher up, till the cosmic ones, it is very hard, if not in fact
impossible, to encounter among all the immensity of the number of objects we can observe two
perfectly identical ones. \\
On the other hand, on the atomic and on the quantum scales we make the assumption that, say,
every two electrons are absolutely the same from whatever point of view relevant to physics. \\

Well, existence on the quantum or atomic scales must therefore be just about ... infinitely
boring ... \\

Or perhaps, not ... \\
In which case the various entities of the quantum realms may actually have features relevant
physically, yet so far not known to us, and thus not taken into account. \\

Therefore, one possible way to see the randomness in a variety of quantum phenomena is as an
effect of the combination of the "operative clumsiness", mentioned in section 4, with the
"conceptual clumsiness" pointed to above, a clumsiness manifested in our present view of
seeing existence on the quantum scales so endlessly boring due to the immense number of
entities which are assumed to be perfectly identical from whatever physically relevant point
of view ... \\

Indeed, once the "conceptual clumsiness" of that "boringly repetitive" structure of the
quantum realms is set aside, it is conceivable that, say for instance, electrons or photons do
in fact spread across a considerable spectrum with respect to a number of yet not known
physical features. And then, the randomness we may observe with respect to them may be the
result of the statistical distribution of such features, features which in ways not yet known
make electrons, photons, and so on, behave somewhat differently during our operatively and
conceptually clumsy present kind of experiments. \\ \\

{\bf 6. Let Us Have Some Mixture ...} \\

In Stapp, several recent proposals relating to a quantum theory of consciousness are mentioned.
One of them is due to Hameroff \& Penrose, in which three of the presently fascinating ideas
are mixed together in a somewhat ad-hoc manner, in the hope of producing an effect in which
the resulting whole may be larger than the sum of its parts. The respective parts are the
celebrated G\"{o}del incompleteness result of the early 1930s, the recent discovery of the
microtubular structure of the neurons, and some ideas from the ongoing and still far from
conclusive research related to quantum gravity. \\

A problem with such mixes of ideas is that, so often, they do not really fit together in order
to form a more organic whole. What Hameroff \& Penrose use in the hope of constituting a more
genuine whole are some suggested estimates of typical time scales of the action of gravitation
within the brain which, being of the order of tenth of a second, seem to correlate with those
associated with conscious processes. \\
The validity and relevance of such estimates is, however, an open question, given that the
possible actions of gravity on brain, mind, neurological processes, or thinking, let alone the
respective more precise ways of manifestation, are still highly hypothetical. \\ \\

{\bf 7. The Alleged Dualism of Descartes} \\

It is fashionable to label Descartes a "dualist", or even "substance dualist", Stapp. What is
missed is a more thorough understanding of the world-views of thinkers in Europe of those
times. To mention a few of them, Copernicus, Kepler, Galileo, Pascal, Descartes, Newton,
Leibniz, or Spinoza were deeply religious men in the Judaeo-Christian tradition. Consequently,
none of them - and this includes Descartes as well - could possibly be anything else but
fervent "monists". \\

As for "dualism", or for that matter, "substance dualism", chemistry is practicing it without
any objections from any quarter, and it does so in a most successful manner, when it divides
itself into its "inorganic" and "organic" branches. \\

Biology proceeds in a yet more dramatic manner, when it makes an essential differentiation
between "living organisms" and all other forms of matter. And such a differentiation is by no
means arbitrary or superficial. For instance, only plants are able to turn through their
metabolism inorganic, thus clearly non-living matter, into living one. And by far most of the
plants only use inorganic matter in their metabolic processes. Animals, on the other hand,
must use in their metabolism mostly plants or other animals, since they cannot live only on
inorganic intake. \\

As for Descartes, his division in "res cogitans" and "res extensa" was in his own view but of
course about the two branches of a unique tree, two branches which grow out from the same one
and only, universal and all encompassing, eternal grace of God's act of creation. In this way,
what is labelled as mere "dualism" is in the case of Descartes but about the two surface
aspects of manifestation of the fundamental "monism" underlying, creating, and for evermore
sustaining them. \\

In this regard, unless seen in ways similar with Descartes and his famous fellow thinkers of
those older times, the so called "mind-body" or "brain-mind" problems are but problems of
relating together two branches cut off a living tree, and with the tree lost, forgotten, or
denied to exist, or even to have ever existed ... \\

But as at the beginning of section 1, let us stop again for a moment, and take one or more
steps back, in order to gain some perspective. \\
Indeed, we can use that typically human ability of consciousness which allows it to be {\it
self-referential}. \\

And then we can ask the question : \\

\begin{quote}

When one thinks about solving a duality, like for instance, that of the "brain-mind" problem,
is it not that such a thinking happens under the aim of non-dualism, happens under the sign of
an intended monism ?

\end{quote}

Therefore, either that unique tree from which the two terms of duality branched out has ever
existed or not, we, by trying to solve the problem of that duality do in fact plant a tree, a
unique tree in our own thinking, hoping that the two terms of duality may somehow be grafted
into it in an organic manner ... \\
This is, indeed, what we in fact do - consciously or not - when trying to go beyond any
duality ... \\

Here however, there seems to be quite some discrepancy between what the "... right hand and
the left hand ..." end up doing. \\

For instance, Stapp lists as the first main objection to dualism its failure to provide some
understanding about the ways the two dual, and so essentially different, terms do interact.
The second main objection, Stapp, is that, in the case of the "mind-body" problem, for
instance, the physical description already gives a causally and deterministically complete
account of what is going on. Thus the mental element of the duality is only left as a sort of
"ghost in the machine", with the physical side being perfectly able to get along all on its
own. \\

The suggested solution, Stapp, of "quantum interactive dualism" is claimed to evade neatly
both of these major objections. Namely, the interaction between the mental and physical is
claimed to be given by the celebrated von Neumann account of the measurement process. The
second main objection above is done away with due to the major and essential difference
between classical, and on the other hand, quantum physical processes. Indeed, the latter ones
- according to the Copenhagen Interpretation - are not causally complete, when taken all alone
and only within themselves. And then it is precisely the mental processes which are alleged to
come and complete the causal structure, and on top of that, they also underlie the structural
relationship between the elements in our streams of conscious experiences. \\

What is missed in such and other similar arguments is that, before everything else, it is the
unique, consistent, and persistent "tree" of the thinking of the respective physicist
concerned with such a duality which does hopefully bring together in an organic manner the two
essentially different "branches". And the "seed" of that "tree" is there - and {\it must} be
there - already in the physicist's thinking before one or another claimed method, such as for
instance von Neumann's measurement theory, is brought in to overcome duality. \\

Unfortunately however, to the extent that we may not be sufficiently aware of the typically
human ability of {\it self-referentiality} of consciousness, we easily tend to miss on that
"tree" in us ...
And we may also miss, Rosinger [2], on a far more fundamental "tree" ...

\end{document}